# Enhancement of hole injection for nitride-based light-emitting devices

S. M. Komirenko and K. W. Kim

*Department of Electrical and Computer Engineering, NC State University, Raleigh, North Carolina 27695;*

V.A. Kochelap

*Institute of Semiconductor Physics, National Academy of Sciences of Ukraine, Kiev-28, 252650, Ukraine;*

J.M. Zavada

*US Army Research Office, Research Triangle Park, North Carolina 27709*

*Abstract:* A novel device design is proposed for a strong enhancement of hole injection current in nitride-based light-emitting heterostructures. Preliminary calculations show orders of magnitude increase in injected hole current when using the proposed superlattice hole injector device based on the real-space transfer concept.

*Introduction:* Further development of group-III-nitride based optoelectronics requires solution of a crucially important problem - obtaining highly doped *p*-type regions, or more precisely, high-density hole currents. It is well established that the difficulties in achieving high hole concentrations originate from: (a) the relatively low solubility of the typical acceptors (Mg, Zn, C, etc.), (b) neutralization of these acceptors by formation of complexes with hydrogen and other material defects, and (c) most importantly, the deep level of known acceptors (about 250 meV for Mg in GaN [1]). Fabrication of heavily p-doped device regions is even more difficult to achieve for AlGaN materials, where energies of the accepter levels are found to be larger [1,2]. To overcome the low acceptor activation problem, it was suggested [3] that a p-doped ternary compound material with a spatially modulated chemical composition [e.g., a superlattice (SL)] can enhance the average hole concentration**.** Calculations [3] and Hall measurements [4,5] support the idea of improved acceptor efficiency in Mg-doped ternary SLs: the average hole concentration can be increased up to one order of magnitude. However the main drawback of this approach is that the most of the holes ionized from the acceptors are localized inside the quantum wells (QWs), which have potential barriers as high as 100 to 400 meV. These barriers hinder participation of the holes in vertical transport required in typical light-emitting devices.

***Proposed approach:*** To increase the overbarrier hole concentration and the vertical hole current, we propose a two-terminal hole injector schematically illustrated in Fig. 1(a).

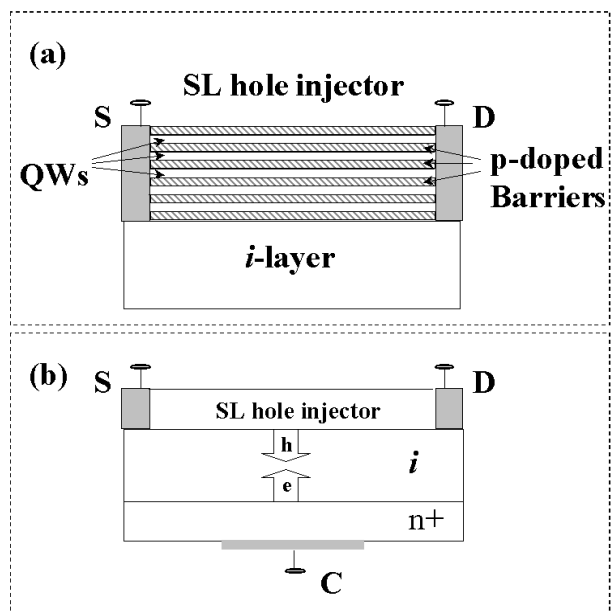

Fig. 1 Schematic illustration for (a) the two terminal SL hole injector and (b) a light-emitting device with the SL hole injector.

The injector consists of a *p*-doped SL base and two contacts S and D. The injector is separated from the rest of the device by an *i*-region. A bias voltage applied between the S and D contacts provides *lateral* hole acceleration and increases the effective temperature of the holes $T_h$, resulting in an enhancement of overbarrier hot-hole concentration. This is known as the real-space transfer effect [6].

In general, a light-emitting device can be thought of as a *three terminal device,* schematically shown in Fig. 1(b), with a hole-injector region, an intrinsic barrier *i*-layer, and an *n*-doped region, having a contact C. If a *p*-doped SL is used as a hot-hole injector, the device can operate as a charge injection transistor [7]. With the D contact as ground and a positive voltage $V_S$ applied to the S contact, the injector would yield a lateral hole current and heating of the holes. Assuming that a negative bias $V_C$ is applied to the cathode C, then both the hot holes from the SL and the electrons from the n-region would be injected into the *i*-layer, as illustrated in Fig. 1(b). Emission of light would occur in the *i*-layer as the result of electron-hole recombination.

*Parameter estimates:* Since the holes in the nitrides have a large effective mass [1], we can assume that several subbands are populated in the QWs composing the SL. Then, the estimate for the concentration of the holes moving freely overbarrier is: $n_h = n_s \exp(-U_b/k_B T_h)/L_{QW}$. Here, $U_b$ is the barrier height, $k_B$ is the Boltzmann constant, $L_{QW}$ is the thickness of the QWs, $n_s$ is the two-dimensional hole concentration in the QWs related to the average hole concentration $n_{av}$ in the SL by the relation $n_s = n_{av}(L_B + L_{QW})$, where $L_B$ is the thickness of the barriers. The hole temperature can be found from the energy balance equation: $e\mu_h E^2 = k_B[T_h - T_l]/\tau(\varepsilon)$; where $e$ is the elementary charge, $\mu_h$ is the *lateral* mobility of the holes in the QWs, $E$ is the *lateral* electric field, $T_l$ is the lattice temperature and $\tau(\varepsilon)$ is the energy relaxation time for the confined holes.

As a specific example, we consider a GaN/Al$_x$Ga$_{1-x}$N SL with $L_{QW}=L_B$. Let us assume that the barrier layers are selectively doped by Mg providing an average hole concentration of $10^{18}$ cm$^{-3}$ (this level of acceptor activation has been reported in different experimental papers [4,5]). For the Al fraction x=0.2, the value of $U_b$ can be found in range 100 - 300 meV, depending on the degree of accuracy to which the effects of modulation of the valence band by polarization fields are (or can be) taken into account [4]. Setting $U_b$ = 200 meV, one can estimate that under equilibrium, at $T_h=T_l$=300 K, almost all ionized holes are confined into the QWs, the concentration of the holes moving freely over barrier is $n_h \approx 8.6 \times 10^{14}$ cm$^{-3}$.

To estimate the effect of hole heating by a lateral electric field, the magnitude of the *lateral* hole mobility should be estimated. It is known that in the nitride semiconductors the hole mobility is low: $\mu_h \leq 150$ cm$^2$/Vs was measured in the best bulk samples [8]. The hole mobilities measured in the doped SLs are reported to be less than 20 cm$^2$/Vs. However, the latter values should be regarded as the hole mobility for the *vertical* transport and the lateral mobility has not yet been measured. It is expected that the lateral mobility is appreciably larger (especially in the modulation doped SLs) and, at least, comparable to the values of bulk mobility. This behavior is well known for other III-V and Si/SiGe heterostructures [9]. Thus, we may assume that in high-quality SL structures the lateral hole mobilities can reach the values of $\mu_h \approx$ 50 - 150 cm$^2$/Vs. Estimating the hole energy relaxation time to be on the order of $10^{-11}$ s$^{-1}$ [10], we obtain that even under very moderate lateral electric fields of 3 – 5 kV/cm, a hole temperature of about 450 K can be achieved. This would increase the overbarrier hole concentration above $10^{16}$ cm$^{-3}$, i.e., by more than an order of magnitude [$n_h(T_h=450K)/n_h(T_h=300K)>13$]. It is easy to see that the hole concentration enhancement can reach several orders of magnitude when a higher electric field, a larger barrier height, or a lower lattice temperature is used.

Assuming a reasonable source-drain bias range of 1-3 V, we obtain that the fields necessary to effectively elevate the hole temperature can be achieved for SLs with the lateral dimensions of the order of 2 – 10 µm. Scaling down the intercontact distance between the S and D contacts in the injector will lead to an increase in the hole injection current and will permit reduction in the lateral bias. A large area hot-hole injector can be achieved by fabricating multiple contacts to the SL in a sequence of S-D-S-D...

*Conclusion:* Lateral electric current through a p-doped group-III-nitride SL can appreciably enhance the hole concentration participating in the vertical transport. Such a SL with properly designed source and drain contacts providing hole heating can serve as a micron-scale hot-hole injector for nitride-based light emitting devices. In addition to the enhancement of the vertical electric current in the device, the hot-hole injector can perform an important controlling function by modulating, for example, the current through the active region and subsequent light emission. Planar design of the proposed structure allows integration of the injectors into a microchip and, consequently, permits an increase in the light emitting area.

*Acknowledgment:* This work was supported in part by the US Army Research Office.

**References**

1   PEARTON, S. J., ZOLPER, J. C., SHUI, R. J., and REN, R. J.: 'GaN: Processing, defects, and devices', J. Appl. Phys., 1999, 86, pp. 1-78

2   TANAKA, T., WATANABE, A., AMANO, H., KOBAYASHI, Y., AKASAKI, I., YAMAZAKI, S., and KOIKE, M.: 'p-type conduction in Mg-doped GaN and $Al_{0.08}Ga_{0.92}N$ grown by metalorganic vapor phase epitaxy', Appl. Phys. Lett., 1994, 65, pp. 593-595

3   SCHUBERT, E. F., GRIESHABERT, W., and GOEPFERT, I. D.: 'Enhancement of deep acceptor activation in semiconductors by superlattice doping', Appl. Phys. Lett., 1996, 69, pp. 3737-3739

4   KOZODOY, P., HANSEN, M., DENBAARS, S. P., and MISHRA, U.: 'Enhanced Mg doped efficiency in $Al_{0.2}Ga_{0.8}N$/GaN superlattices', Appl. Phys. Lett., 1999, 74, pp. 3681-3683

5   SHEU, J. K., CHI, G. C., and JOU, M. J.: 'Low operational voltage of InGaN/GaN light-emitting diodes by using a Mg-doped $Al_{0.15}Ga_{0.85}N$/GaN superlattice', Electron. Lett., 2001, 22, pp. 160-162

6   GRIBNIKOV, Z. S., HESS, K., and KOSINSKY, G. A.: 'Nonlocal and nonlinear transport in semiconductors - real space-transfer effects', J. Appl. Phys., 1995, 77, pp. 1337-1373

7   LURYI, S.: 'Light-emitting devices based on the real-space transfer of hot electrons', Appl. Phys. Lett., 1991, 58, pp. 1727-1729

8   GASKILL, D. K., ROWLAND, L. B., and DOVERSPIKE, K.: 'Electrical transport properties of AlN, GaN and AlGaN', in *Properties of Group III Nitrides* (ed. Edgar, J.), EMIS Datareviews Series, 1995, No. 11, pp. 101-116

9   MITIN, V. V., KOCHELAP, V. A., and STROSCIO, M. A.: 'Quantum Heterostructures' (Cambridge University Press, New York, 1999)

10  OGUZMAN, I. H., KOLNÍK, J., BRENNAN, K. F., WANG, R., FANG, T.-N., and RUDEN, P. P.: 'Hole transport properties of bulk zinc-blende and wurtzite phases of GaN based on an ensemble Monte Carlo calculation including a full zone band structure', J. Appl. Phys., 1996, 80, pp. 4429-4436